\documentclass[a4paper,11pt]{article}
\pdfoutput=1 

\usepackage{jcappub} 

\usepackage[T1]{fontenc} 
\usepackage{amssymb,amsmath,amsfonts,amsbsy,graphicx}
\usepackage{color}
\usepackage{psfrag}
\usepackage{stackrel}
\usepackage{bm}
\usepackage{tikz}
\usetikzlibrary{positioning}
\usepackage{siunitx}
\usetikzlibrary{intersections}
\usepackage{url}
\usepackage{svg}
\usepackage{comment}
\usepackage{here}

\newcommand\lcdm{$\Lambda$CDM }

\newcommand{\Msun}{M_\odot}

\newcommand{\orcid}[1]{$\,$\href{https://orcid.org/#1}{\includesvg[width=10pt]{orcid}}}

\newcommand{\Msunhinv}{M_\odot /h}
\newcommand{\vir}{\mathrm{vir}}
\newcommand{\host}{\mathrm{h}}
\newcommand{\sub}{\mathrm{s}}
\newcommand{\myr}{\mathrm{Myr}}

\newcommand{\mr}{\mathrm}
\newcommand{\mrd}{\mathrm{d}}

\title{
\boldmath Survival of Gas in Subhalos and Its Impact on the 21 cm Forest Signals: Insights from Hydrodynamic Simulations
}

\author[a]{Genki Naruse,}
\author[g]{Kenji Hasegawa,}
\author[b,c]{Kenji Kadota,}
\author[d]{Hiroyuki Tashiro,}
\author[a,e,f]{Kiyotomo Ichiki}

\affiliation[a]{
    Graduate School of Science, Department of Physics,\\
    Nagoya University,\\
    Nagoya, 464-8602, Japan
    }
\affiliation[b]{
    School of Fundamental Physics and Mathematical Sciences,\\ Hangzhou Institute for Advanced Study, \\
    University of Chinese Academy of Sciences (HIAS-UCAS),\\
    Hangzhou 310024, China
    }
\affiliation[c]{
    International Centre for Theoretical Physics Asia-Pacific (ICTP-AP),\\
    Beijing/Hangzhou, China
    }
\affiliation[d]{
    Center for Education and Innovation, Sojo University,\\
    Ikeda, Nishi-ku, Kumamoto, 860-0082 Japan
    }
\affiliation[e]{
    Kobayashi-Maskawa Institute for the Origin of Particles and the Universe,\\
    Nagoya University,\\Nagoya, 464-8602, Japan
    }
\affiliation[f]{
    Institute for Advanced Research, Nagoya University,\\
    Furocho, Chikusa-ku, Nagoya, Aichi 464-8602, Japan
    }
\affiliation[g]{
    National Institute of Technology,\\
    Suzuka College,\\
    Shirokocho, Suzuka, Mie 510-0294, Japan
    }

\emailAdd{naruse.genki.v5@s.mail.nagoya-u.ac.jp}
\emailAdd{hasegawa-k@mech.suzuka-ct.ac.jp}
\emailAdd{kadota@ucas.ac.cn}
\emailAdd{hiroyuki@ed.sojo-u.ac.jp}
\emailAdd{ichiki.kiyotomo.a9@f.mail.nagoya-u.ac.jp}

\abstract{
Understanding the survival of gas within subhalos under various astrophysical processes is crucial for elucidating cosmic structure formation and evolution. We study the resilience of gas in subhalos, focusing on the impact of tidal and ram pressure stripping through hydrodynamic simulations. Our results uncover significant gas stripping primarily driven by ram pressure effects, which also profoundly influence the gas distribution within these subhalos. Notably, despite their vulnerability to ram pressure effects, the low-mass subhalos can play a pivotal role in influencing the observable characteristics of cosmic structures due to their large abundance.

Specifically, we explore the application of our findings to the 21 cm forest, showing how the survival dynamics of gas in subhalos can modulate the 21 cm optical depth, a key probe for detecting minihalos in the pre-reionization era. 
Our previous study demonstrated that the 21-cm optical depth can be enhanced by the subhalos, but the effects of tidal and ram pressure stripping on the subhalo abundance have not been fully considered. 
In this work, we further investigate the contribution of subhalos to the 21 cm optical depth with hydrodynamic simulations, particularly highlighting the trajectories and fates of subhalos within mass ranges of \(10^{4-6} M_{\odot}h^{-1}\) in a host halo of \(10^7 M_{\odot}h^{-1}\), and subhalos within mass range of \(10^{4-5} M_{\odot}h^{-1}\) in a host halo of \(10^6 M_{\odot}h^{-1}\). 
Despite their susceptibility to ram pressure stripping, the contribution of abundant low-mass subhalos to the 21-cm optical depth is more significant than that of their massive counterparts primarily due to their greater abundance. We find that the 21-cm optical depth can be increased by a factor of approximately two due to the abundant low-mass subhalos. However, this enhancement is about twice as low as previously estimated in our earlier study, a discrepancy attributed to the effects of ram pressure stripping. Our work provides critical insights into the gas dynamics within subhalos in the early universe, highlighting their resilience against environmental stripping effects, and their impact on observable 21-cm signals.
}

\begin{document}
\maketitle
\flushbottom

\section{Introduction}\label{sec:intro}

In the standard \lcdm model, the structure formation proceeds according to the bottom-up fashion, and minihalos whose CDM mass ranges from $\sim 10^4\Msun$ to $\sim 10^7 \Msun$ are expected to collapse before the reionization epoch. 
Therefore, discovering such small-scale objects is important to confirm the CDM cosmology. 
However, minihalos have not been observed yet, even with the latest telescopes, because their mass range is too small to detect. 

Since minihalos are expected to contain abundant HI gas, observing HI is one way to detect the minihalos. 
Unfortunately, the well-known HI Lyman-$\alpha$ forest is not available for this purpose because its cross-section is too large, and the absorption feature is easily damped by HI gas in the intergalactic medium (IGM). 
An alternative absorption signal from HI gas is the hyperfine transition, the so-called 21 cm line, whose cross-section is very small compared to Lyman-$\alpha$. 
Though the 21 cm emission from minihalos could be strong~\cite{2002ApJ...572L.123I}, it is difficult to spatially resolve minihalos~\cite{YajimaLi2014,Tanaka2018}. 
On the other hand, observing the 21 cm absorption feature is still a promising approach to detect minihalos. 
Previous theoretical studies have shown that the 21 cm optical depth of minihalos is $\sim 0.01-0.1$~\cite{2020PhRvD.102b3522S, 2020PhRvD.101d3516S}, hence the Square Kilometre Array (SKA) would detect the signal if there are radio loud background sources which are brighter than $\sim 10$ mJy.  

According to the bottom-up scenario, minihalos universally contain the internal dense structures so-called subhalos.
Recently, Kadota et al. (2023)~\cite{2023JCAP...03..017K} (hereafter K23) studied the impact of subhalos to the 21 cm optical depth. 
The virial temperature $T_\vir$ is related to the halo mass as $T_\vir \propto M^{2/3}$. Hence, the subhalos have lower temperatures than the host halo. 
For the sake of the fact that 21 cm optical depth is roughly proportional to the inverse of the spin temperature \cite{2001PhR...349..125B, 2002ApJ...579....1F,2011MNRAS.410.2025X, 2011MNRAS.417.1480M}, which is close to the gas kinetic temperature in a halo, the subhalos likely enhance the 21 cm optical depth. 
In K23, we analytically estimated the 21 cm optical depth of subhalos in a host halo and concluded that 
the larger host halo can have the larger boost of 21 cm optical depth by subhalos because the larger host halo has the lower 21 cm optical depth and the larger number of subhalos inside. For example, the subhalos can boost the total 21 optical depth by a factor of 4 in the case of $10^7 \Msunhinv$ host halo. 
However, some significant effects were not taken into account in K23.  
For instance, when subhalos move in a host halo, they always suffer from tidal and ram pressure forces. 
Subhalos are consequently expected to lose their gas mass during their motion, reducing their optical depth. 
It is also expected that the dynamical friction effectively works in a gas-rich system because gas particles have pressure and tend to accumulate behind an orbiting object, resulting in the enhancement of the drag force, and subhalos could quickly fall towards the center of the host halo, where the tidal and ram pressure forces strongly work.  
In addition to these dynamical phenomena, there are thermal aspects to consider, including compressional heating. 
In K23, the halos are assumed to be in the isothermal state at all times. But, in the realistic case, the compression would heat the subhalos and decrease their optical depth.  

In order to obtain a more accurate estimation of the subhalo contribution to the 21 cm forest signal, it is necessary to take into account these physical effects. However, these effects are difficult to consider analytically. Therefore, in this study, we conduct hydrodynamic simulations to evaluate the 21 cm optical depth while accounting for these effects.
To compute the 21 cm optical depth, taking into account subhalos, we take two steps. 
In the first step, we perform hydrodynamic simulations in which a subhalo of various masses moves within a host halo.
Using simulation data, we evaluate the time evolution and spatial distribution of 21 cm optical depth originating from a moving subhalo.
Then, in the second step, we calculate the total 21 cm optical depth with subhalos following a subhalo mass function by performing the subhalo mass function weighted integration of the optical depths of subhalos derived from the simulations.  
Hereafter, the subscript $\host$ denotes the physical value of a host halo, and $\sub$ does that of subhalos.

Throughout this study, we assume the \lcdm model with cosmological parameters: $h = 0.674$, $\Omega_{m} = 0.315$, $\Omega_b h^2 = 0.02237$, $\Omega_{\Lambda}=0.685$ based on Plank 2018 result \cite{2020A&A...641A...6P}.
Our paper is organized as follows. 
In Section \ref{sec:singlehalo}, we explain the setup of our simulations and show the simulation results. 
In Section \ref{sec:minihalotau}, we describe how to compute and estimate the 21 cm optical depth of the subhalos along the line of sight (LOS) inside the minihalo using our simulation data, and present our results and compare them to those of K23. 
Finally, Sections \ref{sec:discussion} and \ref{sec:Summary} are respectively devoted to the discussion and summary.

\section{Hydrodynamic simulations}
\label{sec:singlehalo}
As mentioned in the introduction, to evaluate the 21 cm optical depth of a minihalo, we need to know the dynamical and thermal evolution of subhalos as they move inside the host halo. 
For this purpose, we conduct numerical simulations in which dark matter (DM) and baryon dynamics are solved simultaneously.
We utilize a hydrodynamic simulation code \textit{START}, \cite{2010MNRAS.407.2632H} which employs smoothed particle hydrodynamics (SPH) method.  
The halo masses we consider in this work are less than $10^7M_\odot$ where atomic cooling hardly works. 
We also ignore hydrogen molecular cooling, which is usually effective in minihalos, for simplicity. We discuss this point later in Section \ref{sec:discussion}. 

\subsection{Simulation Setup}
\label{sec:setup}
We employ the same initial condition used in K23 to directly compare results (Appendix \ref{sec:haloprof} describes the initial conditions concretely).  
A host halo and a subhalo have the same density profile except for the region outside the virial radius. 
We first place a DM halo with the Navarro, Frenk \& White~(NFW) profile~\cite{2000ApJ...540...39A, 1997ApJ...490..493N} at $z=10$ for which the concentration parameter is taken from the fitting formula of~\cite{2021MNRAS.506.4210I}. 
Then, we distribute the baryon component to obey the isothermal hydrostatic equilibrium with $T_{\rm gas}=T_{\rm vir}$. 
With this setup, the halo artificially expands due to the vacuum boundary condition. 
Therefore, we also distribute the IGM particles surrounding the host halo to moderate the artificial expansion of the halo. 
As for the IGM particles, we extrapolate the gas 
density profile up to $2R_{\vir, \host}$
and set $T = 2.18 \rm K$ which corresponds to the temperature of adiabatically cooled gas at $z=10$. 
The enclosed gas mass fraction is set to be $f_\mr{b} = \Omega_\mr{b}/\Omega_\mr{m} = 0.156$. 
In this work, we employ the host halo masses ($M_{\host}$) of $10^7 M_\odot/h$ and $10^6 M_\odot/h$. 
In each run, a host halo contains a subhalo with $m_{\sub}=10^6$, $5 \times 10^5$, $10^5$, $5\times10^4$, $3\times10^4$, or $10^4 \Msunhinv$. 
We set the gas-particle mass to $10 f_\mr{b} \Msunhinv$ and the DM particle mass to $10(1 - f_\mr{b}) \Msunhinv$ so that the mass resolution is identical in all runs. 
The simulations are performed for 3 times a characteristic dynamical time scale ($3\sqrt{1/G \bar{\rho}_\host} \approx 150 \myr$), 
which roughly corresponds to the timescale that the host halo can keep its current mass without merges at $z=10$. (see explanation \ref{sec:appsimtime} for more details.)

\begin{figure}[tbp]
\centering
\includegraphics[width=0.5\textwidth]{figures/initial_setup_warrow.pdf}
\caption{
An initial state of a minihalo in the equatorial ($x-y)$ plane. 
The color indicates the gas temperature. The region where the temperature is higher than $\sim10^3\rm K$ corresponds to the minihalo with the size of $R_{\vir,\host}$, and the outer low-temperature region is the IGM region. 
A subhalo is initially at the position of ($x$, $y$, $z$) = ($R_{\vir, \host}$, 0, 0) and has the initial velocity of $|\sqrt{GM_{\vir,\host}}/R_{\vir,\host}|$ with different entry angles of 0  (the magenta arrow), 30 (yellow), 60 (yellow-green), and 90 (cyan) degrees. 
}
\label{fig:simset}
\end{figure}

We show an initial state of the simulation in Fig. \ref{fig:simset}. 
In all runs, a subhalo with $m_{\sub}$ starts to move from the position ($x$, $y$, $z$) = ($R_{\vir, \host}$, 0, 0) (see Fig. \ref{fig:simset}). 
We consider four entry angles, $\theta_{\rm ini}$ of $0^\circ$ (circular motion), $30^\circ$, $60^\circ$, and $90^\circ$ (radial motion). 
It is expected that the potential energy and the kinetic energy almost balance with each other when subhalos merge with the host halo. 
Therefore, though a velocity dispersion exists in reality, we suppose an averaged velocity is the Kepler velocity and such velocity dispersion does not affect our conclusions. 
We fix the magnitude of the initial velocity to be the Kepler velocity $v_{\rm ini}=\sqrt{GM_{\host}/R_{\vir,\host}} = 9.39 \rm km/s$. 
We also assume that the initial velocity is independent of the entry angle for simplicity.

\subsection{Evolution of Subhalos}
\label{sec:subevlo}

\begin{figure}[tbp]
\centering
\includegraphics[width=1.0\linewidth]{figures/temp_timeevolv_Ms1E5_iniv00_lowerhemisphere.pdf}
\hfill
\includegraphics[width=1.0\linewidth]{figures/temp_timeevolv_Ms1E5_iniv30_lowerhemisphere.pdf}
\hfill
\includegraphics[width=1.0\linewidth]{figures/temp_timeevolv_Ms1E4_iniv00_lowerhemisphere.pdf}
\caption{
Two-dimensional color maps showing the time evolution of the gas temperature on the equatorial plane (z=0). Snapshots every 47.2 Myr from the initial time are shown from left to right. 
The top, middle, and bottom rows respectively represent results with the parameter set of $(m_{\sub},\theta_{\rm ini})=(10^5 \Msunhinv, 0^\circ)$, $(10^5 \Msunhinv, 30^\circ)$, and $(10^4 \Msunhinv, 0^\circ)$ in the host halo of $10^7 \Msunhinv$. 
The IGM particles outside host halo fall towards the boundary of the host halo as time passes. However, the density profile inside the host halo hardly changes and does not aggravate our results in these simulation times. 
}
\label{fig:Tevl}
\end{figure}

\begin{figure}[tbp]
\centering
\includegraphics[scale=0.7]{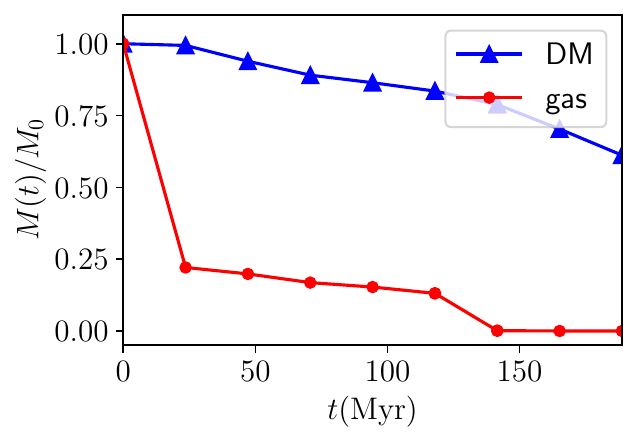}
\hfill
\includegraphics[scale=0.7]{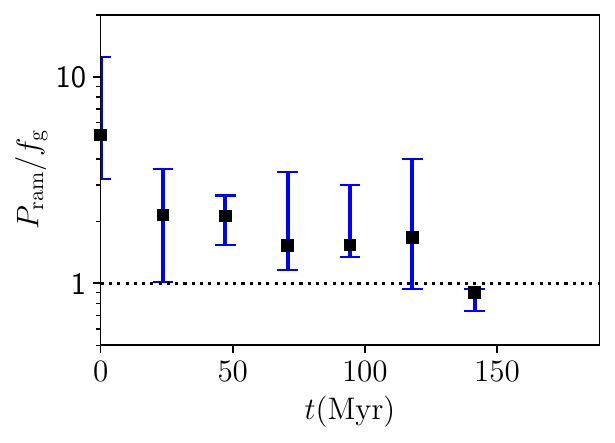}
\caption{Both panels are the results of $(m_{\sub},\theta_{\rm ini})=(10^5 \Msunhinv, 0^\circ)$ in the host halo of $10^7 \Msunhinv$. 
Left: The mass fractions of gas (red circles) and DM (blue triangles) to the initial gas/DM mass ($M_0$) of the subhalo as a function of time.
After $90$ \% of the initial mass is removed, the halo is considered a disrupted halo in this analysis.
Right: The ratio of the ram pressure $P_\mr{ram}$ to the maximum gravitational restoring force per unit area $f_\mr{g}$ as a function of time \cite{2008MNRAS.383..593M}. 
The $10$ \% of the outermost bounded particles are used to evaluate $f_g$ for this analysis,
while all bounded particles are used when the number of the bound particles is less than $100$. 
The squares represent the median values, and the error bars are drawn from the maximum and minimum values.   
}
\label{fig:ramp}
\end{figure}

We first show the evolution of subhalos in a host halo of $10^7 \Msunhinv$ in Fig. \ref{fig:Tevl}. From left to right, each panel shows the snapshot at $t \approx 0$, 47.2 $\myr$, 94.4 $\myr$ and 142 $\myr$, respectively. 
In the figures, the color represents gas temperature. 
The top row shows the result of the parameter set of $(m_{\sub},\theta_{\rm ini})=(10^5 \Msunhinv, 0^\circ)$. 
In this case, the subhalo is initially heated up to $1.5 T_{\vir}$ by the compressional heating. 
After that, the gas in the subhalo is considerably stripped off during $\sim 100 \myr$. 
The left panel of Fig. \ref{fig:ramp} quantitatively displays the time evolution of the gas and DM mass fractions, defined as the fraction of the bound gas and DM mass to each initial mass.
For this analysis, we select particles bound to the subhalo's potential by comparing the potential energy and kinetic energy of particles. Hereafter we simply call those particles "bound" particles. 
By $\sim 100 \myr$, $20\%$ of the DM is stripped off from the subhalo. On the other hand, the gas mass stripping is stronger than the DM mass stripping, and only $10\%$ of the initial gas remains.  
Such a difference is caused by the forces they receive. 
In fact, the DM component is only affected by the tidal force, while the gas component is affected by both the tidal force and the ram pressure. 
Therefore, the ram pressure effect mainly causes the gas mass stripping. 
To gain a better understanding of ram pressure stripping, the right panel in Fig. \ref{fig:ramp} shows the time evolution of the ratio between the ram pressure ($P_\mr{ram}$) and the maximum gravitational restoring force per unit area ($f_\mr{g}$) \cite{2008MNRAS.383..593M} expressed as 
\begin{equation}
    P_\mr{ram} / f_\mr{g} = \frac{\rho_{\mr{gas}, \host} v_\mr{orb}^2}{\gamma G m_\sub (R_\mr{s}) \rho_{\mr{gas}, \sub}(R_\mr{s}) / R_\mr{s}}~, 
    \label{eq:pg}
\end{equation}
where $\gamma$ is a parameter depending on the density profile. We apply $\gamma = \pi/2$ in this study because the singular isothermal sphere can approximate the gas density profile (appendix \ref{sec:appgasdist}). 
Eq. (\ref{eq:pg}) indicates that $f_{\rm g}$ is initially proportional to $m_{\rm s}^{2/3}$ and less massive halos are more susceptible to the ram pressure than massive subhalos.  
Figure \ref{fig:ramp} shows that the ram pressure considerably overcomes the gravitational restoring force per area at the initial phase, resulting in significant gas mass loss. 
Since the subhalo density approximately follows the singular isothermal profile, the gravitational restoring force is stronger at the inner part of the subhalo as $f_{\rm g}\propto \rho^2 R_\sub^2 \propto R_\sub^{-2}$. 
Therefore, as the outer side is stripped by ram pressure stripping, $f_{\rm g}$ eventually approaches $P_{\rm ram}$, mitigating mass loss in later stages.  
This is the common trend in all simulation runs.

The middle row of Fig. \ref{fig:Tevl} shows the evolution of $(m_{\sub},\theta_{\rm ini})=(10^5 \Msunhinv, 30^\circ)$. 
Since the subhalo passes through the inner high-density region, the ram pressure strongly works and the gas stripping is more prominent than the case of $(m_{\sub},\theta_{\rm ini})=(10^5 \Msunhinv, 0^\circ)$ (the top row). 
Finally, the bottom row of Fig. \ref{fig:Tevl} represents the case of a lower subhalo mass of $m_{\sub}=10^4 \Msunhinv$. 
In this case, ram pressure significantly affects the subhalo, leading to a drastic stripping of gas due to a weak gravitational restoring force.

\begin{figure}[tbp]
\centering
\includegraphics[scale=0.7]{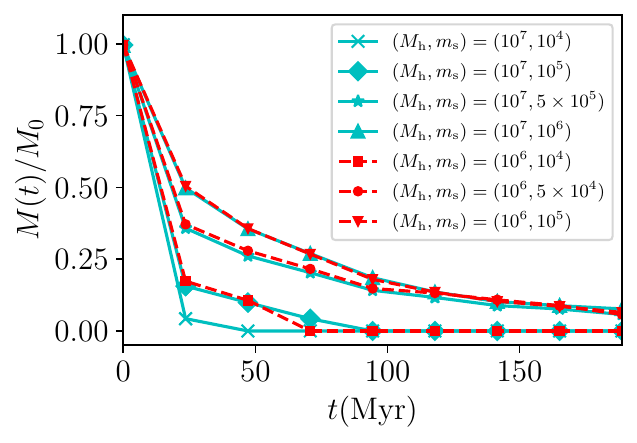}
\hfill
\includegraphics[scale=0.5]{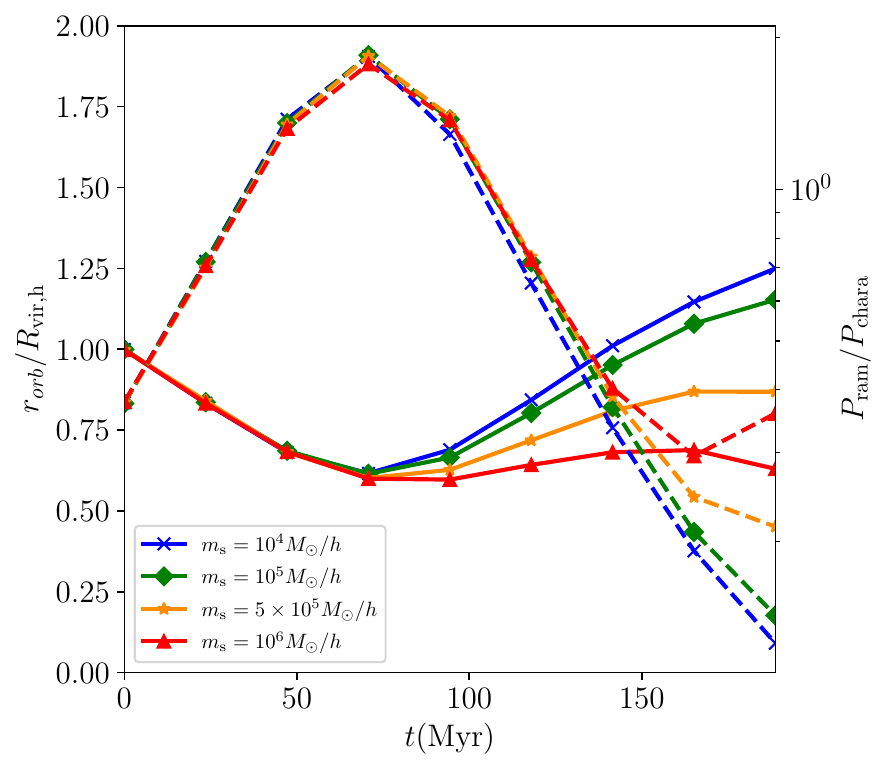}
\caption{
Left: The time evolution of the gas mass fractions of subhalos 
in the host halo of $M_\host = 10^7 \Msunhinv$ (cyan) and $10^6 \Msunhinv$ (red). The represented subhalos in each host halo are $(M_\host, m_\sub) = (10^7, 10^4)$ (crosses), $(10^7, 10^5)$ (diamonds), $(10^7, 5 \times 10^5)$ (stars), $(10^7, 10^6)$ (triangles), $(10^6, 10^4)$ (squares), $(10^6, 5 \times 10^4)$ (circles), and $(10^6, 10^5)$ (inversed triangles) (omitting the unit $\Msunhinv$). 
The initial entry angle is $30^\circ$ for all cases. 
Right: The time evolution of the subhalo's orbital radius $r_\mr{orb}$ (the solid lines) and the ram pressure (the dashed lines) in the host halo of $10^7 \Msunhinv$. 
The orbital radius and the ram pressure are respectively normalized by the host halo virial radius and by the characteristic value of $P_\mr{chara} = \bar{\rho}_\mr{halo, b}\; v_\mr{cir}(R_{\vir, \host})^2$, where $\bar{\rho}_\mr{halo, b}$ is the averaged baryon density of a virialized halo collapsed at $z=10$ 
and $v_\mr{cir}(R_{\vir, \host})$ is the circular velocity at the virial radius.
}
\label{fig:massfrac}
\end{figure}

The left panel of Fig. \ref{fig:massfrac} summarizes the evolution of the baryon mass fractions for various subhalo masses with $\theta_{\rm ini}=30^\circ$. 
As discussed above, the balance between the ram pressure and the gravitational restoring force determines the final mass fraction. 
As stated above, the less massive subhalos have less gravitational restoring force per area and more susceptible to the ram pressure for a given ram pressure. 
Therefore, fixing host halo mass, the final mass fraction tends to be smaller for the less massive subhalos. 
However, at a later phase of $t\gtrsim 75 \myr$, the mass dependence seems to be relatively weak. 
The evolution of the orbital radii and the resultant ram pressure of subhalos in the host halo of $10^7 \Msunhinv$ is shown in the right panel of Fig. \ref{fig:massfrac} to understand the reason why the mass dependence becomes weak. 
Compared to less massive subhalos, massive subhalos' orbital radii tend to be small at $t\gtrsim 75 \myr$ due to the dynamical friction. 
Consequently, the ram pressure strongly works for massive subhalos. Then, we focus on a comparison between the results of the $10^7 \Msunhinv$ host halo and the $10^6 \Msunhinv$ host halo. 
We should note that the evolution of the mass fraction is controlled by $P_\mr{ram}/ f_\mr{g}$ which is initially proportional to $(M_\host/ m_\sub)^{2/3}$ (see Eq. (\ref{eq:pg})). 
Indeed, you can find that each evolution of the mass fractions is very similar when the mass ratio $M_\host/ m_\sub$ is the same. 
When we fix the subhalo mass, $P_\mr{ram}/ f_\mr{g}$ is larger for more massive host halo. 
Therefore, the ram pressure stripping of a subhalo with a given $m_\sub$ is more noticeable in $10^7 \Msunhinv$ host halo.

In conclusion, we find that the gas temperature of subhalos is initially increased by a factor of about 1.3-1.5 due to the compressional heating, and then a large amount of the gas is stripped by the ram pressure.
The effect of the ram pressure is noticeable for less massive subhalos that are destroyed by $\sim 100 \myr$. 
Both the heating and the disruption would lead to a reduction in the 21 cm optical depth. 
Consequently, the contribution, especially by smaller subhalos in the larger host halo, declines compared to K23. 
We should note that, even after most of the gas is stripped from a subhalo, the stripped gas still keeps a lower gas temperature than the host halo gas, as shown in Fig. \ref{fig:Tevl}. 
Thus, the disrupted subhalos still contribute to the total 21 cm optical, as we will show later.

\section{Calculating 21 cm optical depth}
\label{sec:minihalotau}
In this section, we estimate the optical depths originating from subhalos using the hydrodynamic simulation results.  
The 21 cm optical depth per impact parameter from the center of the host halo $\alpha_\host$ is calculated as
\begin{equation}
    \label{tauonesub:eq}
    \tau (\alpha_\host) = \frac{3 h_\mathrm{p} c^3 A_{10}}{32 \pi k_\mathrm{B} \nu_{21}^2}\int_{-l_\mathrm{max}\left(\alpha_\host\right)}^{l_\mathrm{max}\left(\alpha_\host\right)} \mathrm{d}l 
    \frac{n_\mathrm{HI}\left( R_\host \right)}{T_\mathrm{s}\left( R_\host \right) \sqrt{\pi} b(R_\host)}, 
\end{equation}
where $h_{\rm p}$, $c$, $A_{10}$, $k_{\rm B}$, and $\nu_{21}$ are the Planck constant, the speed of light, the Einstein coefficient, the Boltzmann constant, and the frequency corresponding to 21 cm line, respectively. 
$l$ is the path along the LOS with $l = \sqrt{R_{\host}^2 - \alpha_\host^2}$, $l_\mr{max} = \sqrt{R_{\vir, \host}^2 - \alpha_\host^2}$, and $b$ is the velocity dispersion $b^2 = 2 k_\mr{B} T(R_\host)/m_\mr{p}$. 
The spin temperature $T_{\rm s}$ is computed using the collisional coupling \cite{1969ApJ...158..423A, 2005ApJ...622.1356Z, 2006ApJ...637L...1K, 2023PASJ...75S..33V}. 

\begin{figure}[tbp]
\centering
\includegraphics[width=0.9\linewidth]{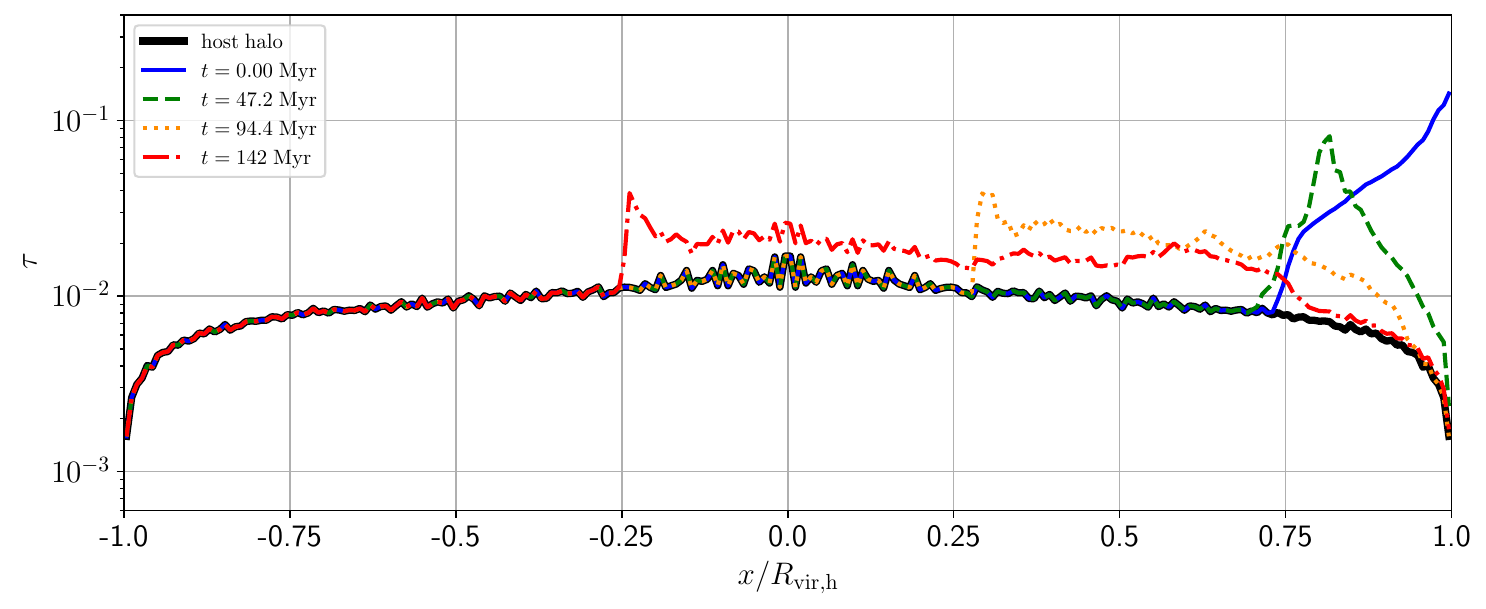}
 \caption{The optical depth along the LOS parallel to the y-axis in the equatorial plane (z = 0) for $(m_{\sub}, \theta_{\rm ini})=(10^5 \Msunhinv, 0^\circ)$ in the host halo of $10^7 \Msunhinv$, 
 which corresponds to the top row of Fig. \ref{fig:Tevl}. 
 It is noted that the horizontal axis corresponds to the $x$-coordinate in Fig. \ref{fig:Tevl} but is normalized by the virial radius of the host halo. 
 The thick black curve represents the optical depth without subhalos, while the thin curves the optical depths with a subhalo at $t=0.00~\myr$ (solid, blue), $t=47.2~\myr$ (dashed green), $t=94.4~\myr$ (dotted orange), $t=142~\myr$ (dash-dotted red). 
 When $t=142~\myr$, over $90$ \% of the subhalo gas mass is stripped off.
 }
\label{fig:subtau_equat}
\end{figure}

Figure \ref{fig:subtau_equat} shows the spatial distribution at different times of the optical depth of a minihalo containing only one subhalo with $(m_{\sub},\theta_{\rm ini})=(10^5 \Msunhinv,0^\circ)$ along the LOS parallel to the y-axis in the equatorial plane. 
The horizontal axis is the coordinate from the center. 
A clear peak initially appears at $x/ R_{\vir, \host} = 1.0$,  because the subhalo gas is not stripped yet. 
As time passes, the peak gradually decreases due to the gas stripping and almost disappears by $t \approx 140 \myr$. 
However, as mentioned above, the stripped gas has a lower temperature than the host halo, so the subhalo slightly contributes to the total optical depth. 

\begin{figure}[tbp]
\centering
\includegraphics[width=0.5\textwidth]{figures/cartoon_ave.pdf}
\caption{ 
A schematic view for getting $\tau_\sub^x$ as a function of $\alpha_\host$. We take an average of $\tau_\sub^x$ in the thin ring between $\alpha_\host$ and $\alpha_\host + \mrd \alpha_\host$ on the $y$-$z$ plane (represented by red shaded region). As for $\tau_\sub^y$ and $\tau_\sub^z$, we consider $x$-$z$ plane and $x$-$y$ plane respectively. 
}
\label{fig:comptau}
\end{figure}

We then evaluate the total 21 cm optical depth of subhalos which follow a subhalo mass function $\frac{\mrd n_\sub}{\mrd m_\sub}$. 
Here, we use the same subhalo mass function as used in K23 \cite{2023JCAP...03..017K, 2019Galax...7...68A, 2018PhRvD..97l3002H}, but use a different spatial distribution of subhalos from K23 in which they assumed that the number density of subhalos obeys NFW profile or uniform distribution inside a host halo. 
In our simulations, the position of a subhalo depends on the elapsed time and its entry angle. 
Assuming subhalos randomly accrete onto the host halo,  
we take the time average over our total simulation time for a given impact parameter $\alpha_\host$ and get the time-averaged value $\bar{\tau}$. (The total time for our simulations is about $3 \sqrt{1/G \bar{\rho}_\host}$ (3 times a characteristic dynamical time scale) as mentioned in Section \ref{sec:setup}.)

Eventually, the total optical depth of subhalos for a given entry angle $\theta_{\rm ini}$ is expressed as 
\begin{equation}
    \label{tausub:eq}
    \tau_\mathrm{s}^\omega \left(\alpha_\host, \theta_{\rm ini} \right) = \int_{10^4 \Msunhinv} \mrd m_\sub 
    \left. \frac{\mrd n_\sub}{\mrd m_\sub}\left(m_\sub \right) \right|_{M_\host = M_\host'} \bar{\tau} (m_\sub,\alpha_\host, \theta_{\rm ini}),  
\end{equation}
where $M_\host'$ is the value of the host halo mass we employed in the simulation run: $10^7 \Msunhinv$ or $10^6 \Msunhinv$. 
Here, as mentioned above, this $\tau_\mr{s}$ relies on the direction of LOS which is represented by the subscript $\omega(=x,y,z)$. We shall discuss this in the next paragraph. 
In addition, it is noted that we took the average of the optical depth of the subhalo with the given $\alpha_\host$ for $\tau_\mathrm{s}^\omega$ as a function of a given $\alpha_\host$. Fig. \ref{fig:comptau} shows a complementary illustration of $\tau_\mathrm{s}^\omega(\alpha_\host)$ as an example of $\omega = x$. 
The applied subhalo mass function contains a cut-off above which the number of subhalos exponentially decrease. 
The lower limit of $10^4 \Msunhinv$ roughly corresponds to the Jeans mass. 
With respect to the mass integration, we only have the discrete dataset ranging from $10^4 \Msunhinv$ to $10^6 \Msunhinv$. 
Therefore, we interpolate and extrapolate them linearly with 60 mass bins to complete the integration using SciPy.
 
\begin{figure}[tbp]
\centering
\includegraphics[width=0.9\linewidth]{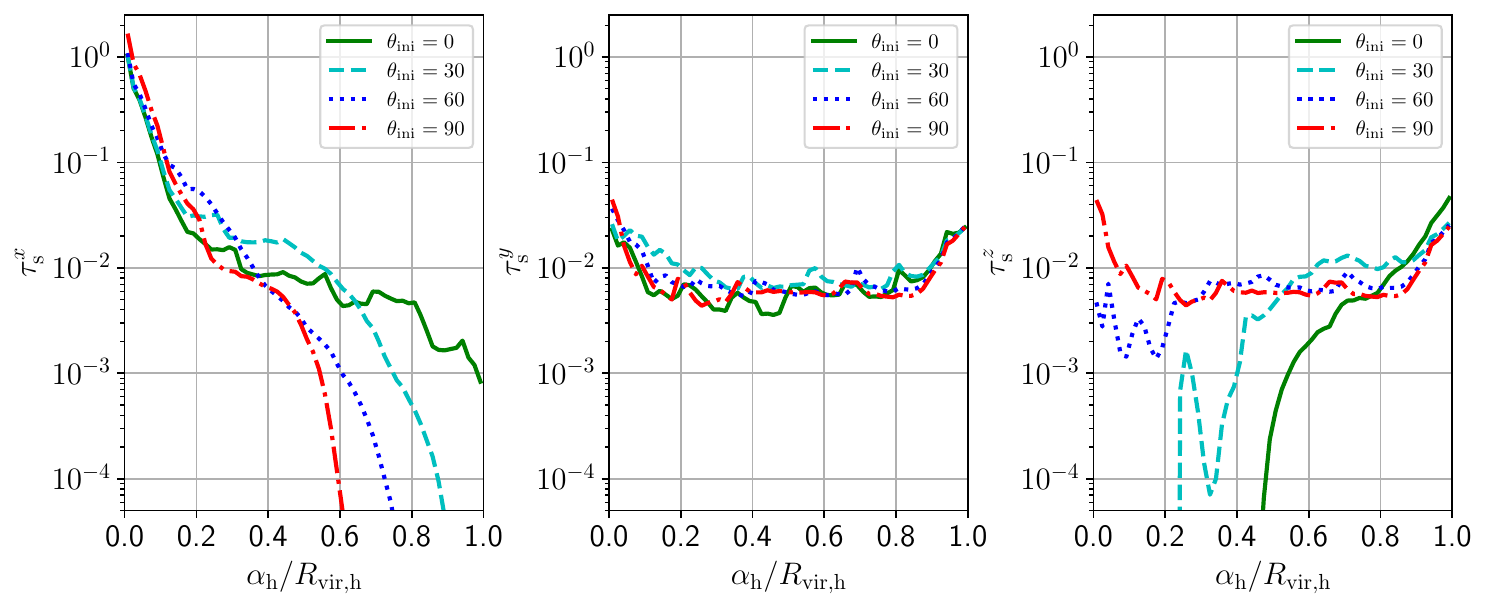}
\caption{The time-averaged optical depth of subhalos following the subhalo mass function in the host halo of $10^7 \Msunhinv$. 
The shown optical depths are averaged in a thin ring with a given impact parameter. 
The solid green, dashed cyan, dotted blue, and dash-dotted red curves show the results with $\theta_{\rm ini}=0^\circ$, $\theta_{\rm ini}=30^\circ$, $\theta_{\rm ini}=60^\circ$, and $\theta_{\rm ini}=90^\circ$, respectively.  
The left, central, and right panels, respectively, show the values along the x-, y-, and z-axes (cf. Fig. \ref{fig:simset}). 
}
\label{fig:subtaulos}
\end{figure}

Figure \ref{fig:subtaulos} shows $\tau_\mathrm{s}^\omega (\alpha_\host, \theta_{\rm ini})$ for several entry angles. 
The left panel of Fig. \ref{fig:subtaulos} represents the results with the LOS parallel to the x-axis (cf. Fig. \ref{fig:Tevl} and \ref{fig:comptau}).
Since the subhalo initially exists at $\alpha_{\host}=0$ without disruption, the optical depth at $\alpha_{\host}\sim 0$ is very high for all entry angles. 
When $\theta_{\rm ini}=0^\circ$, the subhalo is moving in a circular orbit, and hence the gas in the subhalo is extended up to $R_{\vir,\host}$. 
The central panel shows the case with the LOS being along the $y$-axis. 
Obviously, the distribution of the optical depth hardly depends on $\theta_{\rm ini}$ in this case.  
The high optical depth at the outer region simply comes from the subhalos at the initial stage. 
The optical depth at the central region is also relatively high because all subhalos inevitably pass through the $\alpha_{\host}\approx0$ region. 
The right panel for the LOS is parallel to the $z$-axis. 
Similar to the case of the central panel, the optical depth in the outer region is high for all entry angles owing to the subhalos at the initial position. 
The subhalos with $\theta_{\rm ini}=90^\circ$ straightly fall towards the center, while those with $\theta_{\rm ini}=0^\circ$ never reach the central region. 
Therefore, the optical depth in the inner region depends strongly on the entry angle. 

Finally, we compare our results with the analytic estimation by K23. 
For this purpose, we take an average of $\tau_{\sub}^\omega$ over three line-of-sights and four entry angles, 
\begin{equation}
    \label{tauave:eq}
    \tau_{\mr{ave},\sub} (\alpha_\host) = \frac{\sum_{\omega} \sum_{\theta_{\rm ini}} 
    \tau_{\sub}^\omega \left(\alpha_\host, \theta_{\rm ini} \right)}{n_\omega n_{\theta_\mr{ini}}}.
\end{equation}

\begin{figure}[tbp]
\centering    
\includegraphics[width=0.9\linewidth]{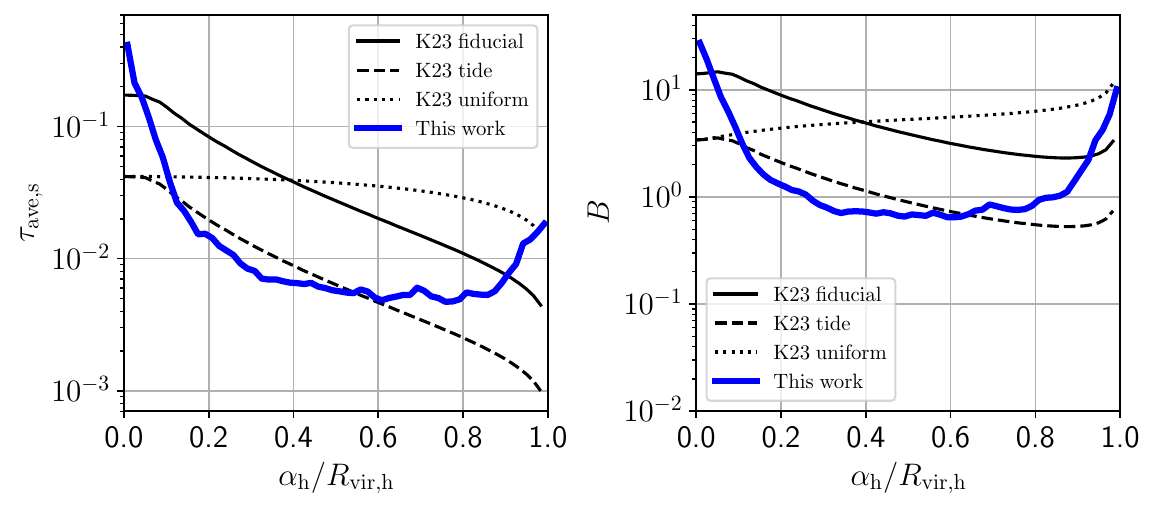}
\caption{The subhalo optical depth (Left) and the boost factor $B$ (Right) in the host halo of $10^7 \Msunhinv$
as a function of the impact parameter normalized to the host halo virial radius at z = 10. 
The thick (blue) curve is our result, and the thin (black) curves are the results from K23. 
As for K23 results, the solid curve represents the case where the subhalos are not affected by the stripping and are distributed following the NFW profile of the host halo. 
The dashed curve represents the case that the outer regions $r > 0.77 r_{\sub, \sub}$ of the subhalos are assumed to be stripped completely, where $r_{\sub, \sub}$ is the subhalo's scale radius. 
The dotted curve represents the case that the subhalos without mass stripping are uniformly distributed. 
}
\label{fig:subtau2}
\vfill\vspace{3mm}
\includegraphics[width=0.9\linewidth]{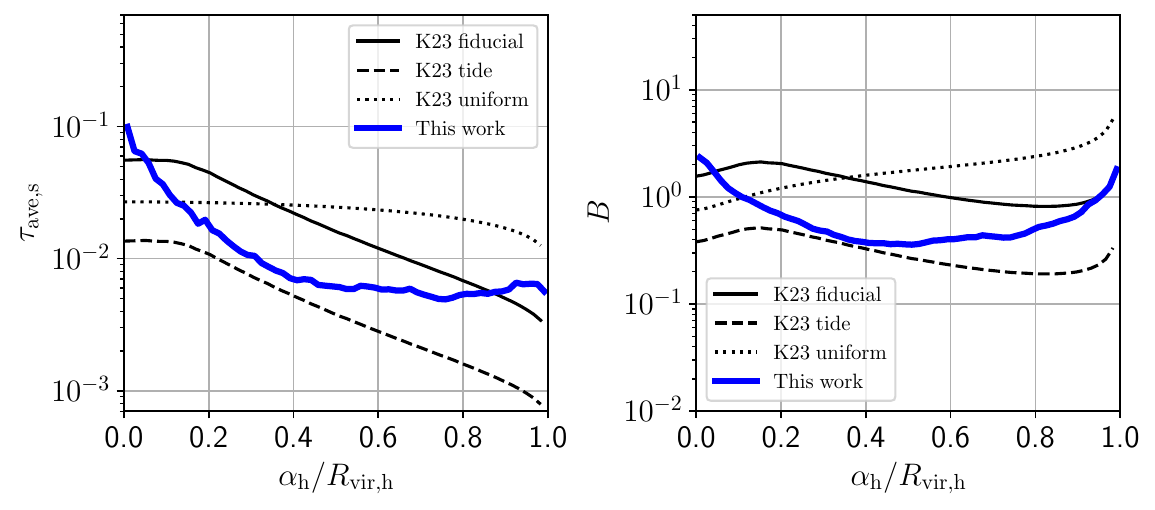}
\caption{
The subhalo optical depth(Left) and the boost factor $B$ (Right) as a function of the impact parameter in the same way as Fig \ref{fig:subtau2} but in the host halo of $10^6 \Msunhinv$. 
}
\label{fig:subtau_mh1e6}
\end{figure}

The denominator represents the number of the ensemble of our simulation. The profile of $\tau_{\rm ave, \sub}$ would depend on LOS directions taken in the ensemble, and we for simplicity take the average over LOS directions parallel to x, y, and z axes in our study shown in Fig. \ref{fig:subtaulos}. (In this study, $n_\omega = 3$ and $n_{\theta_\mr{ini}} = 4$, so the denominator is $n_\omega n_{\theta_\mr{ini}} = 12$.) 
We also quantify the subhalos' contribution by using the boost factor, defined as the subhalo optical depth ratio to the host halo optical depth, $B \equiv \tau_{\rm ave. \sub} / \tau_\host$ as is in K23. 
Figure \ref{fig:subtau2} and Figure \ref{fig:subtau_mh1e6} show the optical depth of subhalos (left) and the boost factor (right), in the host halo of $10^7 \Msunhinv$ and $10^6 \Msunhinv$ respectively. 
In each panel, the thick (blue) curve is our simulation result, and the thin (black) ones are results from K23 \footnote{K23 gave the conservative estimates by neglecting the contributions from subhalos which do not totally fit inside the host halo virial radius when the subhalo centers are at the outer edge of a host halo. In our figures, for easier comparison with our study, the K23 results have been slightly modified to include the contributions from all of those subhalos whose centers are inside the host radius.}.

With respect to the outer region of $\alpha_\host/R_{\vir,\host} \gtrsim 0.9$, our result shows higher optical depth than the K23 "fiducial." 
This is caused by the difference in the spatial distribution of subhalos. 
The NFW profile assumed in the K23 "fiducial" model causes the number density of subhalos at this region to be smaller than our simulations.
On the other hand, the distribution of subhalos in our simulations roughly corresponds to the uniform distribution because the optical depth shown here is the averaged value. 
In fact, the optical depth at $\alpha_\host /R_{\vir,\host} \gtrsim 0.9$ in the simulation is almost consistent with that in the K23 "uniform" model. 
In the middle region of $0.2 \lesssim \alpha_\host /R_{\vir,\host} < 0.9$, our result is lower than the K23 "fiducial" model because the gas mass stripping by the ram pressure works and the subhalos contribute to the optical depth modestly. 
Finally, with regard to the innermost region $\alpha_\host/R_{\vir,\host} < 0.2$, the optical depth steeply increases towards the center and eventually exceeds the K23 "fiducial" model at $\alpha_\host \approx 0$. This behavior is essentially caused by the same reason as shown in Fig. \ref{fig:subtaulos}. All subhalos pass through the $\alpha_\host \approx 0$ region, except for cases where LOS is parallel to the z-axis (the right panel of \ref{fig:subtaulos}).
In particular, when we chose the LOS along the x-axis (the left panel of Fig. \ref{fig:subtaulos}), all subhalos are initially at $\alpha_\host \approx 0$ without destruction. 
Therefore, we notice that this profile of $\tau_\mr{ave,s}$ should be modified if we consider the ensemble average over all LOS directions. 

We emphasize that the most important value here is the optical depth averaged over the apparent area which does not depend on the choice of the LOS since the spatial distribution of $\tau_{\rm ave, \sub}$ in each minihalo cannot be observationally resolved even with the SKA. 
The boost factor of a minihalo averaged over the apparent area is given by 
\begin{equation}
    \label{eq:tauareaave}
    B_\mr{halo} = \frac{\int^{R_{\vir, \host}}_{0} 2 \pi \alpha B(\alpha) \mrd \alpha}{\int^{R_{\vir, \host}}_{0} 2 \pi \alpha \mrd \alpha}. 
\end{equation}
where $B$ is defined below Eq. (\ref{tauave:eq}).
In the case of $10^7 \Msunhinv$ host halo, we find that the boost factor derived from our simulations is $B_\mr{halo, sim} \approx 1.67$, 
whereas that in the K23 "fiducial" model is $B_\mr{halo, K23} \approx 3.67$. 
Thus, we conclude that the hydrodynamic effects, such as the ram pressure and the compressional heating, reduce the 21 cm optical depth of subhalos to down to less than a half ($\sim 45 \%$) compared to the analytic estimate. 
It should be emphasized that subhalos in a minihalo whose mass is $10^7 \Msunhinv$ can enhance the optical depth by more than a factor $(1+B_\mr{halo}) \sim 2.5$, even if the hydrodynamic effects are considered. 
On the other hand, in the case of the $10^6 \Msunhinv$ host halo, we find $B_\mr{halo,sim} \approx 0.603$ whereas $B_{\mr{halo, K23}} \approx 1.10$ since the hydrodynamic effect reduces the 21 cm optical depths of subhalos down to $\sim 55 \%$ compared to the analytic estimate. 

Comparing these results, it is confirmed that the boost by the subhalos is larger for the $10^7 \Msunhinv$ host halo and this is qualitatively consistent with the previous analytic estimate by K23.  
In K23, the reason for this trend is simply due to the low optical depth of host halo with $M_\host=10^7 \Msunhinv$ and abundant subhalos in the massive host halo. 
On the other hand, in the simulations, the destruction of subhalos by the ram pressure is more significant in a massive host halo as shown by Fig.\ref{fig:massfrac}. 
Therefore, the ratio between the boost factors of $M_\host=10^6 \Msunhinv$ host halo and $M_\host=10^7 \Msunhinv$ host halo is larger than that estimated by K23, i.e., $0.603/1.67 \approx 0.361$ for the simulation and $1.10/3.67\approx 0.300$. 

\begin{figure}[tb]
\centering
\includegraphics[width=0.9\linewidth]{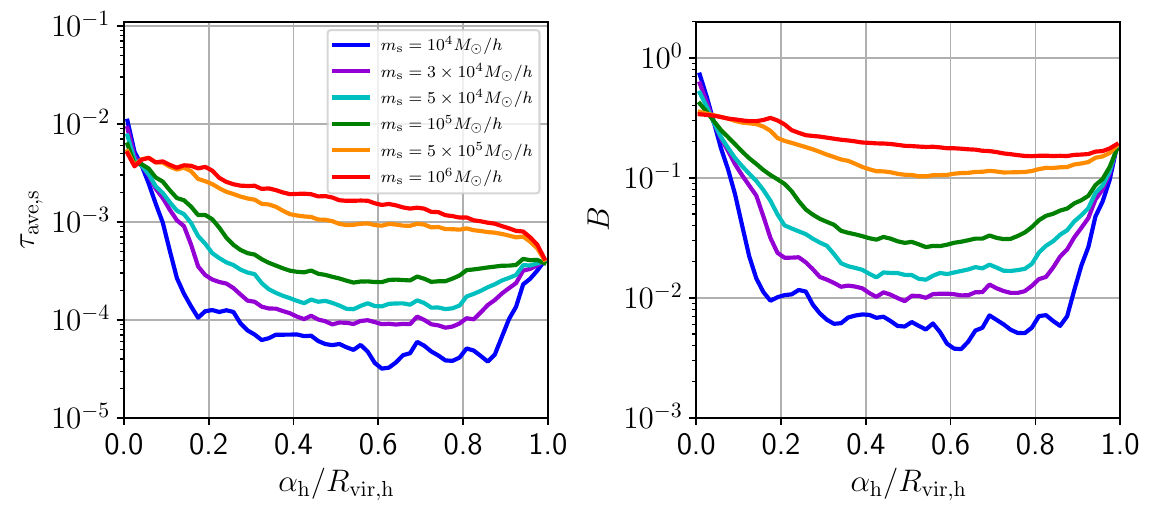}
\caption{Left panel: the optical depths of one subhalo with $m_{\sub}=10^4 \Msunhinv$ (blue), $3\times 10^4 \Msunhinv$ (purple), $5\times 10^4 \Msunhinv$ (cyan), $10^5 \Msunhinv$ (green), $5\times 10^5 \Msunhinv$ (orange), and $10^6 \Msunhinv$ (red) in the host halo of $10^7 \Msunhinv$. 
Right panel: the corresponding boost factors. }
\label{fig:massdepend1}
\end{figure}
\begin{figure}[tb]
\centering
\includegraphics[width=0.9\linewidth]{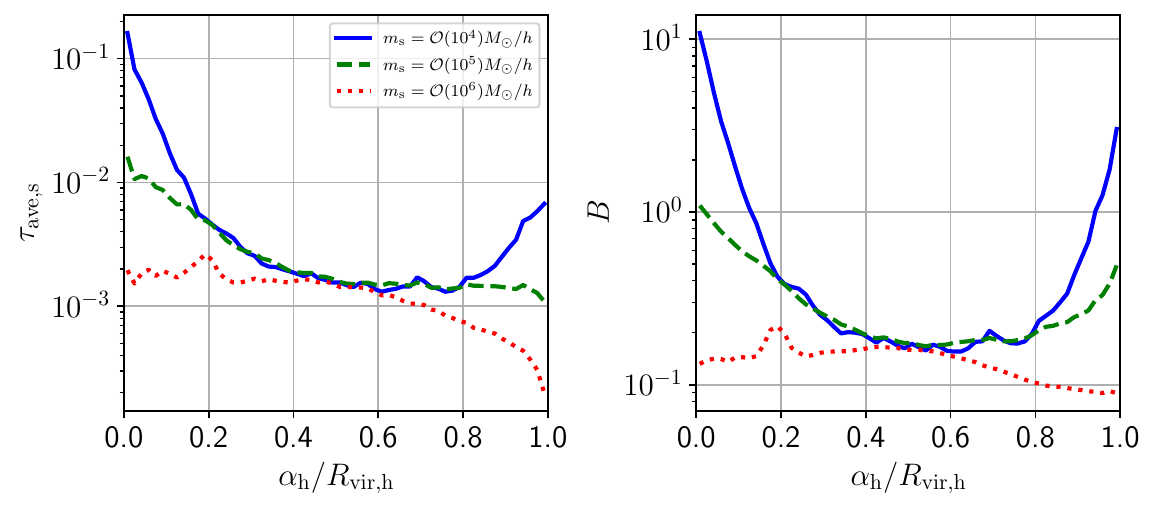}
\caption{Left: The subhalo mass function weighted optical depths ($\tau_\mr{ave}$ in Eq. \ref{tauave:eq}) for three integration regions of $10^4 \Msunhinv \le m_{\sub}<10^5 \Msunhinv$ (solid, blue), $10^5 \Msunhinv \le m_{\sub}<10^6 \Msunhinv$ (dashed, green), and $10^6 \Msunhinv \le m_{\sub}<10^7 \Msunhinv$ (dotted, red) in the host halo of $10^7 \Msunhinv$. 
Right: The corresponding boost factors. 
}
\label{fig:massdepend2}
\end{figure}

In addition, we investigate which mass scale is responsible for the optical depth. 
Fig. \ref{fig:massdepend1} shows the averaged optical depth of one subhalo in the host halo of $10^7 \Msunhinv$ with various masses (left) and the corresponding boost factor (right) (the $\tau$ value is not integrated but is focused on the specific subhalo mass in Eq. (\ref{tausub:eq})). 
One can observe that the optical depth increases with the mass of subhalos.
As stated in Section \ref{sec:intro}, the optical depth of a low-mass halo is higher than that of a massive halo, as long as the subhalo is not affected by the tidal force and ram pressure.   
However, as shown in Fig. \ref{fig:massfrac}, low-mass subhalos are more sensitive to the ram pressure than massive ones. 
When the ram pressure just starts to work, the strong gas stripping for a less massive subhalo compensates for the high optical depth of the subhalo. 
As a result, the optical depth is almost independent of the subhalo mass at $\alpha/R_{\vir, \host} \sim 1$. 
Less massive subhalos considerably lose their gas component as time passes, and the optical depth is smaller than massive subhalos in $0.2 \lesssim \alpha/R_{\vir,\host} \lesssim 0.9$ region. 

The left panel of Fig.\ref{fig:massdepend2} shows the subhalo mass function weighted optical depth in the host halo of $10^7 \Msunhinv$ for three different integration ranges of $10^4 \Msunhinv \le m_{\sub}<10^5 \Msunhinv$, $10^5 \Msunhinv \le m_{\sub}<10^6 \Msunhinv$, and $10^6 \Msunhinv \le m_{\sub}<10^7 \Msunhinv$ (the $\tau$ value is integrated over the such specific range of subhalo mass in Eq. (\ref{tausub:eq})). 
Interestingly, considering the subhalo mass function, the total contribution from more abundant low-mass subhalos is more dominant than that from less abundant high-mass subhalos owing to the strong mass dependence of the adapted subhalo mass function. 
In summary, although the contribution to the optical depth from a low-mass subhalo is smaller than that from a massive subhalo due to the former's vulnerability to gas stripping, the cumulative contribution from low-mass subhalos becomes dominant overall because of their greater abundance. 
This trend is the same for the result of $10^6 \Msunhinv$ host halo.

\section{Discussion}
\label{sec:discussion}
In this subsection, we discuss some points that may affect our results. 
One may claim that the gas mass stripping can be caused by the Kelvin-Helmholtz (KH) and Rayleigh-Taylor (RT) instabilities. 
If this is the case, the standard SPH method is known to overestimate the effects. 
According to previous studies \cite{2008MNRAS.383..593M, 2000ApJ...538..559M}, the KH time-scale can be estimated by, 
\begin{equation}
    \label{eq:kh}
    t_\mr{KH} = 2.19 \times 10^9 \left( \frac{F}{0.1} \right) \left( \frac{m_\sub}{10^9 \Msun} \right)^{1/7}
  \left( \frac{n_\host}{10^{-4} \; \mr{cm}^{-3}} \right)^{-1} \left(\frac{v_\mr{orb}}{10^3 \; \mr{km/s}} \right)^{-1} \mr{yr}, 
\end{equation}
where $F$ is the baryon fraction of the halos, $m_\sub$ is the subhalo mass, $n_\host$ is the number of the hydrogen atoms in the host halo, and $v_\mr{orb}$ is the velocity of the subhalo with respect to the host halo. 
In our simulations, the estimated KH time-scale is 
$t_\mr{KH} \sim 213~\myr$, which is longer than the timescale of the gas stripping of $\sim 100~\myr$. 
Therefore, we conclude that the gas stripping is not mainly caused by the instabilities, and our standard SPH method would not overestimate the mass loss in our simulations. 

We employed a somewhat simplified model for halos in this study.  
First, we assumed the collapse redshift of $z = 10$ for the host halo and the subhalos. 
However, according to the bottom-up scenario, the subhalos should collapse before the formation time of the host halo. 
In this case, the subhalos have higher gas density and are more resistant to gas stripping, and this causes the optical depth of the subhalos to be larger. 
For example, as for $z = 15$, the subhalo gas particles are bounded about 1.5 - 2.0 times longer than $z = 10$ by our simulation.
On the other hand, as shown by \cite{2014PhRvD..90h3003S}, the halos at higher redshift have less optical depth roughly because of their higher virial temperature at higher redshift leading to reducing their own optical depth. 
Hence, the optical depth of the subhalos is expected not to be influenced so much by offsetting these effects even considering the clumpiness of the subhalos or can be larger to some extent than our results if the survivability works effectively. 

As for the temperature distribution, we assumed that subhalos initially have uniform distribution with their virial temperature.  
However, in the case of adiabatic collapse, halos usually have higher temperatures in the central core region than in the outer region.  
The temperature distribution is more complicated with the cooling effect, which is completely neglected in our simulations. 
Molecular hydrogen cooling is known to be effective in a halo whose mass is greater than about $10^5 \Msunhinv$ and practically the molecular hydrogen cooling time scale $t_\mr{cool}$ calculated by the below equation \cite{2013fgu..book.....L, 1997ApJ...474....1T} is much shorter than the dynamical time,  
\begin{equation}
    \label{eq:tcoolm}
    t_\mr{cool} = 5 \times 10^4 f_{\mr{H_2}} \left( \frac{1+z}{20} \right)^{3} \left( \frac{\Delta_\mr{c}}{200} \right)
  \left( 1 + \frac{10 T_\mr{3}^{7/2}}{60 + T_\mr{3}^{4}} \right)^{-1} \exp\left( \frac{512 \mr{K}}{T} \right) \mr{yr}, 
\end{equation}
where $f_{\mr{H2}}$ is $\mr{H_2}$ fraction and $T_{3} = T/(10^3 \mr{K})$.
Therefore, our assumption is valid only when dissociation of molecular hydrogen by photons is effective. 
We will investigate the optical depth in the more realistic in future work.

\section{Summary}
\label{sec:Summary}
In this study, we first conducted hydrodynamic simulations to explore how the hydrodynamic effects, such as the ram pressure and the compressional heating, affect the evolution of subhalos in a host minihalo. 
We find that the subhalos are heated by the compressional heating at the earlier phase and lose a large amount of the gas component due to the ram pressure. 
The effects of the ram pressure are significant for less massive subhalos: in the case of $M_{\sub}=10^5 \Msunhinv$, 90\% of the gas is stripped off during approximately $100$Myr. 

We then used the subhalo mass function to estimate the total contribution of subhalos to the 21 cm optical depth. 
Upon comparing our findings with the analytic study conducted by K23, we found that the optical depth boost caused by subhalos is reduced by hydrodynamic effects to approximately a half of its original value. However, despite the reduction, subhalos can still more than double the optical depth of the host halo. 
We also find that low-mass subhalos are responsible for the optical depth due to their rich abundance, even though the lower-mass subhalos are more vulnerable to gas stripping by the ram pressure.

\section*{Acknowledgements}
Our simulations were carried out on the cluster installed at Nagoya University. 
This work 
is supported in part by
JST SPRING grant Number JPMJSP2125 (GN), 
the JSPS grant number 21H04467, JST FOREST Program JPMJFR20352935, and the JSPS Core-to-Core Program 
(grant number:JPJSCCA20200002, JPJSCCA20200003)(KI), and the JSPS grant number 21K03533 (HT). 

\appendix
\section{Gas and DM density profiles}
\label{sec:haloprof}
Here we introduce the properties of halos in our simulations that are the same as K23. 
The density profile of a DM halo is set to be the NFW profile \cite{2000ApJ...540...39A, 1997ApJ...490..493N}, 
\begin{equation}
    \label{rhonfw:eq1}
    \rho_{\mathrm{NFW}}(r) = \frac{\rho_0}{\frac{r}{r_\mr{s}} \left( 1 + \frac{r}{r_\mr{s}}\right)^2},
\end{equation}
where $\rho_0$ is
\begin{equation}
    \label{rhonfw:eq2}
    \rho_{0} = \frac{M_{\mathrm{DM}} \; c^3}{4\pi R_{\mathrm{vir}}^3 f(c) }
\end{equation}
with $f(x) = \ln{(1+x)} - x/(1+x)$, $r_\mr{s}$ is the scale radius, and $c \equiv R_\vir / r_\mr{s}$ is the concentration parameter. 
$R_{\mathrm{vir}}$ refers to the virial radius \cite{2001PhR...349..125B}, 
\begin{equation}
    \label{rvir:eq}
    R_{\mathrm{vir}} = 0.784 \left( \frac{M}{10^8 h^{-1} M_{\odot}} \right)^{1/3} \left[\frac{\Omega_m}{\Omega_m^z} \frac{\Delta_c}{18\pi^2}\right]^{-1/3}
    \left( \frac{1+z}{10} \right)^{-1} h^{-1} \;\; \mathrm{kpc}.
\end{equation}

Here, $\Delta_c = 18 \pi^2 + 82d -39d^2$ is the over density of the virialized halo collapsing at the redshift $z$ 
with $d = \Omega_\mr{m}^z - 1$ and $\Omega_m^z = \Omega_m (1+z)^3/(\Omega_m(1+z)^3 + \Omega_\Lambda)$. 

As for the gas density profiles $\rho_{\mathrm{g}}(r)$, it is assumed that the gas is isothermal with $T_{\vir}$ and in the hydrostatic equilibrium \cite{2001PhR...349..125B, 1998ApJ...497..555M,2011MNRAS.410.2025X,1998ApJ...509..544S}, 
\begin{equation}
    \label{rhogas:eq1}
    \ln{\rho_{\mathrm{g}}(r)} = \ln{\rho_{g0}} - \frac{\mu m_\mathrm{p}}{2 k_{\mathrm{B}} T_{\mathrm{vir}}} \left[ v_{\mathrm{esc}}^2(0) - v_{\mathrm{esc}}^2(r) \right], 
\end{equation}
where 
\begin{equation}
    \label{vesc:eq}
    v^2_{\mathrm{esc}}(r) = 2 \int^{\infty}_{r}  \frac{GM(r')}{r'^2} dr' = 2V_\mathrm{c}^2 \frac{f(cx)+cx/(1+cx)}{xf(c)}, 
\end{equation}

\begin{equation}
    \label{vcir:eq}
    V_{c} = \sqrt{\frac{GM}{R_{\mathrm{vir}}}} = 23.4 \left( \frac{M}{10^8 h^{-1} M_{\odot}} \right)^{1/3} \left[\frac{\Omega_m}{\Omega_m^z} \frac{\Delta_c}{18\pi^2}\right]^{1/6} \left( \frac{1+z}{10} \right)^{1/2} \;\; \mathrm{km/s},
\end{equation}

\begin{equation}
    \label{rhogas:eq2}
    \rho_{g0}(z) = \frac{(\Delta_c/3) c^3 e^A}{\int_{0}^{c} (1+t)^{A/t} t^2 dt} \left( \frac{\Omega_b}{\Omega_m} \right) \Omega_m^z \rho_\mr{crit}(z),
\end{equation}
where $A = 2c/f(c)$ and 
\begin{equation}
    \label{tvir:eq}
    T_{\mathrm{vir}} = 1.98 \times 10^4 \left( \frac{\mu}{0.6} \right) \left( \frac{M}{10^8 h^{-1} M_{\odot}} \right)^{2/3}
    \left[\frac{\Omega_m}{\Omega_m^z} \frac{\Delta_c}{18\pi^2}\right]^{1/3} \left(\frac{1+z}{10} \right) \;\; \mathrm{K}.
\end{equation}

These equations (\ref{vesc:eq}), (\ref{vcir:eq}), (\ref{rhogas:eq2}), (\ref{tvir:eq}) refer to the square of the escape velocity, circular velocity, the central gas density and the virial temperature respectively. 

\section{the time length of the simulation}
\label{sec:appsimtime}
We evaluate the time for our simulation considering the host halos can maintain their mass. 
To do this, we utilize a halo growth rate (mass accretion rate) of the dark matter halos best fitted by the result of a cosmological N-body simulation, \textit{GUREFT} \cite{2024MNRAS.530.4868Y}. 

In \cite{2024MNRAS.530.4868Y}, the halo growth rate $\mrd M_\host / \mrd t$ between $6 \lesssim z \lesssim 14$ is best fitted as, 
\begin{equation}
    \label{eq:HGrate1}
    \frac{\mrd M_\host}{\mrd t} (M_\host, z) = \beta (z) \left( \frac{M_\host}{10^{12} M_\odot} E(z) \right)^{\alpha(z)}
\end{equation}
where $E(z) = H(z)/H_0 = \sqrt{\Omega_m (1+z)^3 + \Omega_l}$ is the expansion rate of the flat universe and the other parameters are 

\begin{align}
    \alpha(z) &= 0.858 + 1.554a -1.176a^2, \label{eq:HGrate2} \\
    \log_{10} \beta(z) &= 2.578 - 0.989a -1.545a^2 \label{eq:HGrate3}
\end{align}
with scale factor $a$.
We can regard the sufficient simulation timescale as the time it takes for the host halo to acquire a mass equal to its own. Therefore, we calculate the time scale by $M_\host/(\mrd M_\host/\mrd t)$ and that is $\approx 150 \myr$ for $(z, M_\mr{vir}) = (10, 10^7 \Msunhinv)$ and $\approx 147 \myr$ for $(z, M_\host) = (10, 10^6 \Msunhinv)$. 

\section{the approximation of the gas density distribution}
\label{sec:appgasdist}
The gas density profile introduced Eq.(\ref{rhogas:eq1}) in this study is approximated in a manner similar to the isothermal $\beta$ model as follows \cite{1998ApJ...497..555M}, 
\begin{equation}
  \label{eq:rhogasapprox}
  \rho_\mr{g}(r) = \frac{\rho_\mr{g0} A(b)}{\left[ 1 + (r/r_\mr{eff})^2\right]^{3\beta_\mr{eff}/2}}, 
\end{equation}
where $b = 4c/27f(c)$, $A(b) = -0.178b + 0.982$, $r_\mr{eff} = 0.22r_\mr{s}$, and $\beta_\mr{eff} = 0.9b$. 
Then we can approximately regard the gas density profile in our simulations as the singular isothermal profile, $\rho_\mr{g}(r) \propto r^{-2}$, so that we can adopt $\gamma = \pi /2$ in Eq. (\ref{eq:pg}) and larger size subhalos have the larger maximum gravitational restoring force per unit area.

\bibliography{sample}
\bibliographystyle{JHEP}




\end{document}